\documentclass[twocolumn,showpacs,amsfonts,aps,prl,floatfix]{revtex4}

\usepackage{bm}
\usepackage{graphicx}
\usepackage{amsmath}
\usepackage{mathrsfs}

\usepackage{ulem}
\usepackage{color}

\begin{document}


\title{Charmonium Production with QGP and Hadron Gas Effects at SPS and FAIR}

\author{Baoyi Chen}
\address{Department of Physics, Tianjin University, Tianjin 300072, China}

\begin{abstract}
The production of charmonium in heavy-ion collisions is investigated
based on Boltzmann-type transport model for charmonium evolution and
langevin equation for charm quark evolution.
Charmonium suppression and regeneration in both quark-gluon plasma (QGP)
and hadron phase are considered. 
Charm quarks are far from thermalization, and regeneration of charmonium in QGP and hadron gas 
is neglectable at SPS and FAIR. 
At peripheral collisions, charmonium suppression with hadron gas explains the 
experimental data well. But at central collisions, additional suppression 
from deconfined matter (QGP) is necessary for the data. 
This means there should be QGP produced at central collisions, and no QGP produced 
at peripheral collisions at SPS energy. 
Predictions are also made at FAIR $\sqrt{s_{NN}}=7.7$ GeV Au+Au collisions. 
\end{abstract}


\date{\today}

\maketitle

\section*{I. Introduction}
One goal of heavy ion collisions is to
identify and study the properties of the deconfined matter, called
quark-gluon plasma (QGP).
Heavy-flavor bound states turn out to be valuable probes of this
strong coupling medium produced in relativistic heavy ion collisions.
$J/\psi$ consists of $c$ and $\bar c$ with large binding energy. Its suppression
was suggested as a signature of the deconfined medium due to color Debye screening~\cite{matsui:1986}.
$J/\psi$ nuclear modification factor $R_{AA}$~\cite{Adare:2006ns,Abelev:2009qaa,Arsene:2012uj},
the elliptic flows~\cite{Yang:2013},
and the average
transverse momentum square~\cite{Adare:2006ns,Topilskaya:2003} tend to support the existence of QGP at
the Relativistic Heavy Ion Collider (RHIC) and Large Hadron Collider (LHC).
However, at low colliding energies at the Super Proton Synchrotron (SPS) and the Facility for Antiproton and Ion Research (FAIR), 
situation becomes more complicated, because effects of hadron gas are very important on charmonium production, 
compared with effects of QGP. 
In order to identify this, we need to include all the contributions 
such as the effects of cold nuclear matter from colliding nuclei, quark gluon plasma and hadron gas, 
at these low colliding energies~\cite{Grandchamp:2002wp, xianglei:2005}.
The production of $J/\psi$ from the recombination of charm quark pairs in the fireball should also be
considered~\cite{Du:2015wha, Thews:2006, yan:2006ve, yunpeng:2014}.
The nuclear modification factor is defined as
\begin{align}
R_{AA}= {N_{AA}^{\Psi}\over N_{pp}^{\Psi}\cdot N_{coll}}
\end{align}
The numerator is charmonium yield in nucleus-nucleus collisions with medium modification.
There are three contributions to it: initial production from the primordial nucleon collisions,
$c$ and $\bar c$ recombination in QGP, and $D$ and $\bar D$ recombination in hadron gas. 
It helps to understand the mechanisms of charmonium suppression and
production in the medium through detailed
calculations about each of three parts.
The denominator is charmonium production in proton-proton collisions, scaled by the number of binary collisions.
This is to establish the initial number of charmonium without nuclear matter effects

\section*{II. Medium Evolution}
In order to understand the charmonium evolution in heavy ion collisions, 
both hot medium and charmonium evolution
should be treated dynamically.
At SPS $\sqrt{s_{NN}}=17.3$ GeV Pb+Pb and FAIR $\sqrt{s_{NN}}=7.7$ GeV Au+Au collisions, 
we assume the boost invariant expansion of medium in longitudinal direction,
and focus on the medium evolution at transverse plane~\cite{xianglei:2005, Shen:2012vn}.
we use (2+1) dimensional ideal hydro dynamics with the conservation equation of net baryon density,
\begin{align}
\partial _\mu T^{\mu\nu}=0 \\
\partial _\mu(n_B u^\mu)=0
\end{align}
where $T^{\mu\nu}=(e+p)u^\mu u^\nu-g^{\mu \nu} p$ is energy-momentum tensor, $u^\mu$ is
velocity of fluid cells. $n_B$ is net baryon density.
For the initialization of hydrodynamic equations
at FAIR, the total multiplicity density at mid-rapidity is taken as
$450$~\cite{Chen:2011vva}. 
Both the time scales of medium reaching local equilibrium
at SPS and FAIR are taken $\tau_{0}=1 \mathrm{fm/c}$~\cite{xianglei:2005, Shen:2012vn}. 
For other inputs at SPS, they are the same as in~\cite{xianglei:2005}, except that we abandon
the approximation of entropy cut. That approximation assumes hydrodynamics doesn't work for hot medium below a
certain entropy density $s_{nc}$
and all hydrodynamic variables are set as zero.
This cut-off of hot medium will strongly underestimate effects of hadron gas on charmonium evolution.

\begin{figure}[b]
  \includegraphics[width=0.9\linewidth]{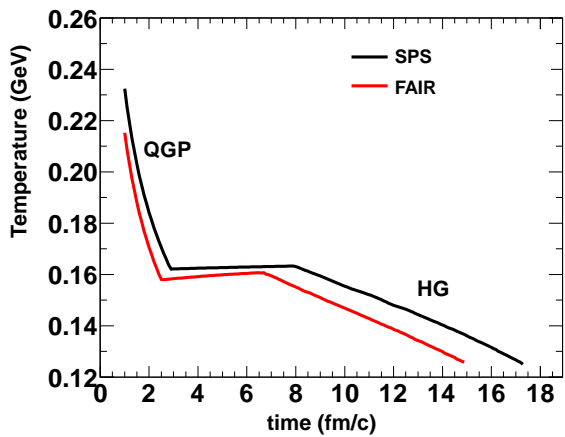}
  \caption{(Color Online) Comparision between temperature evolution at SPS and FAIR. The temperature
is at the center(r=0) of the medium with fixed impact parameter ${\bf b}=0$.}
  \label{fig:1}
\end{figure}

In order to see the difference of hot medium and their effects on charmonium production between FAIR and
SPS, we plot the time evolution of temperature at the center point of the collisions,
see Fig.~\ref{fig:1}. The maximum
temperature of the medium at FAIR is about 215 MeV, which is comparable with
the result in~\cite{Chen:2011vva}.

We assume the first order phase transition between QGP and hadron gas phase. 
The temperature of phase transition $T_c(\mu_B)$ depends on the baryon chemical potential $\mu_B$.
It increases with decreasing $\mu_B$. 
In the mixed phase, medium continues expanding outside which will decrease the net baryon density
and baryon chemical potential. This will increase $T_c$.
That's why even in the mixed phase, the temperature $T_c$ slightly
increases with time instead of being constant.
Also, the net baryon density is larger at FAIR than the value at SPS.
This makes $T_c(\mu_B)$ smaller at FAIR compared with the value at SPS,
just as Fig.~\ref{fig:1} shows.
The initial baryon density with rapidity is taken as
$dN_{B}/ dy=135$ at FAIR~\cite{Chen:2011vva}. 
With the implements of hadron gas contribution
in this work, We are able to discuss the relative contribution of QGP and hadron gas on charmonium 
production at SPS and FAIR. 

\section*{III. Cold Nuclear Matter Effects}
For the initially produced charmonium, they may be dissolved by inelastic collisions
with other nucleons. This suppression exist in pA collisions where QGP should be absent, and is 
called "nucleon absorption", simulated by the inelastic cross-section with nucleons, $\sigma^{abs}$.
$\sigma^{abs}$ is taken as $4.3\ \mathrm{mb}$ for $J/\psi$ and $\chi_c$ ~\cite{xianglei:2005}, and
$7.9$ mb for $\psi^\prime$ at SPS. 
Gluons may also obtain extra energy by multi-scatterings with other nucleons before fusing into a $c\bar c$. 
This effect (called Cronin effect) will change the transverse momentum distribution of
initially produced charmonium~\cite{Gavin:1988tw}, $\langle p_t^2\rangle_{pA}=\langle p_t^2\rangle_{pp}
+a_{gN}\cdot\langle l\rangle$. $\langle l\rangle$ is the average length gluon passes through 
before the reaction $g+g\rightarrow J/\psi +g$, $ a_{gN}$ is taken as $0.076$ $\mathrm{GeV^2/fm}$ 
by fitting the data of pA collisions.
The time scale of producing heavy quark pair is $\sim 1/E_{c\bar c}$, and can be neglected. 
But it takes much longer time for a charm pair to form the charmonium wave function. 
With
uncertainty relation,
$J/\psi$ formation time is about $\tau_{J/\psi}\approx0.35\ \mathrm{fm/c}$ in its local rest frame.
With $\langle p_t^2\rangle_{J/\psi} \sim 1\ \mathrm{(GeV/c)^2}$ at FAIR and SPS,
we can get the Lorentz factor between lab-frame and $J/\psi$ local rest frame. So $J/\psi$ formation time
in lab-frame is about $\tau_{J/\psi}^\mathrm{Lab}=\gamma_{T} \tau_{J/\psi}^\mathrm{LRF}\approx 0.37\ \mathrm{fm/c}$.
It's smaller than $\tau_0$,
which supports the basic assumption that all initially produced
$J/\psi$ are already formed before the transverse expansion of bulk medium.
For excited charmonium states, situation becomes much more complicated and their decay contribution to
the final yield of $J/\psi$ is not dominant.
So we neglect the formation time effect for all charmonium states.
This effect may be important at LHC where $\tau_{J/\psi}^{\mathrm{Lab}}$ is larger and $\tau^\mathrm{LHC}_0$ is smaller.

\section*{IV. QGP and Hadron Gas Effects}
Charmonium mass is large and hard to be thermalized, we use a
Boltzmann-type transport equation~\cite{yan:2006ve,Liu:2010ej,Chen:2012gg,Chen:2013wmr} to describe
the evolution of their phase space distribution in QGP and hadron gas,
\begin{equation}
{\partial f_\Psi\over \partial t} +{\bf v}_\Psi\cdot{\bf \nabla} f_\Psi=
-(\alpha^{\mathrm{QGP}}_\Psi+\alpha^{\mathrm{HG}}_\Psi) f_\Psi 
\label{tran}
\end{equation}
where ${\Psi}$ = ($J/\psi$, $\chi_c$, $\psi^\prime$).
$\alpha_{\Psi}$ for QGP and HG (hadron gas) are
only non-zero in their own phase.
Due to interactions with the hot medium,
charmonium distribution $f_{\Psi}({\bf p},{\bf x},t|{\bf b})$
changes with time which is described by the first term at L.H.S of
equation (\ref{tran}). The second term represents leakage effect, where
charmonia with larger velocity ${\bf v}_\Psi={\bf p}_\Psi/E_\Psi$ tend to
suffer less suppression
by free streaming from the hot medium.
Initially produced charmonium are included through initial distribution at $\tau_0$. 
The cold nuclear matter effects on charmonium
are already discussed above and implemented in the charmonium initial distributions, 
\begin{align}
f({\bf x}_t,{\bf p}_t|{\bf b})&= {\sigma_{pp}^{\Psi}\over \pi}
\int dz_A dz_B \rho_A({\bf x}_t+{{\bf b}\over 2},z_A)\rho_B({\bf x}_t-{{\bf b}\over 2},z_B)\nonumber \\
&\times e^{-\sigma^{abs}(T_A({\bf x}_t+{{\bf b}\over 2} ,z_A,+\infty)
+T_B({\bf x}_t-{{\bf b}\over 2},-\infty,z_B))}\nonumber \\
&\times {1\over \langle p_t^2\rangle} e^{-p_t^2/\langle p_t^2\rangle}
\end{align}
With $\langle p_t^2\rangle= \langle p_t^2\rangle_{pp}+a_{gN}\rho_0^{-1}
(T_A({\bf x}_t+{{\bf b}\over 2},-\infty,z_A) +T_B({\bf x}_t-{{\bf b}\over 2},z_B,\infty))$. 
$T_{A(B)}$ is the thickness function of nucleus A(B), with
$T({\bf x}_t,z_1,z_2)=\int_{z_1}^{z_2}dz\rho({\bf x}_t,z)$. 
Distribution of nucleons $\rho_{A(B)}(x_t,z)$ is taken as Woods-Saxon distribution. 
${\bf b}$ is the impact parameter. 

Terms at R.H.S of equation (\ref{tran}) represents charmonium suppression 
in QGP and HG respectively. For the charmonium suppression in QGP, the reaction is 
$g+J/\psi\leftrightarrow c+{\bar c}$.
The loss term $\alpha_{\Psi}$ is 
\begin{eqnarray}
\label{lg} \alpha^{\mathrm{QGP}}_\Psi({\bf p}_t,{\bf x}_t,\tau|{\bf b}) &=&
{1\over 2E_\Psi}\int{d^3{\bf p}_g\over (2\pi)^3
2E_g}W_{g\Psi}^{c\bar
c}(s)f_g({\bf p}_g,{\bf x}_t,\tau)\nonumber\\
&&\times\ \Theta\left(T({\bf x}_t,\tau|{\bf
b})-T_c\right)
\end{eqnarray}
where $E_g$ and $E_{\Psi}$ are energies of gluon and charmonium respectively. 
The step function $\Theta$ ensures that gluon suppression only happens in QGP.
$f_g= 1/(e^{p_g^\mu u_\mu}-1)$ is the gluon thermal distribution.
$W_{g\Psi}^{c{\bar c}}(s)$ is transition probability of gluon dissociation.
It depends on $s=(p_{\Psi}+p_{g})^2$.
Charmonium survival probability decreases to zero when
approaching to a certain temperature, indicating a dissociation temperature above
which the potential inside $c\bar c$ are totally screened by partons.
The dissociation temperatures for charmonium are $T_d(J/\psi, \chi_c, \psi^\prime)=(2.3, 1.1, 1.1)T_c$
from Schr\"odinger equation with charmonium potential to be its internal energy V=U~\cite{Satz:2005hx}. 

\begin{figure}[b]
  \includegraphics[width=0.9\linewidth]{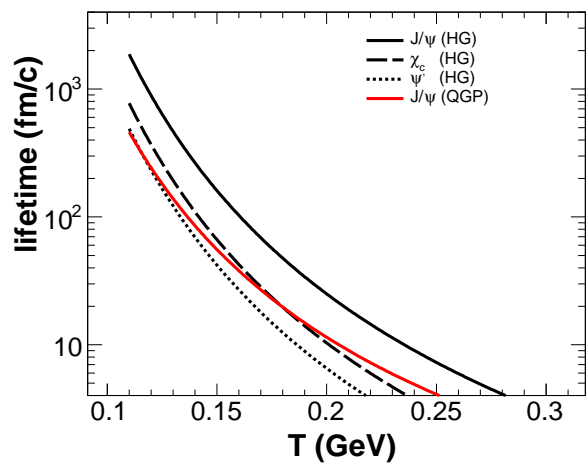}
  \caption{(Color Online) Lifetimes of different charmonium states with temperature in QGP and hadron gas. 
The velocity of medium is taken as $v=0.5$ and 
transverse momentum of charmonium is taken as a typical value $p_\Psi=1$ GeV/c, with the same 
direction of the velocity of medium. 
}
  \label{figlife}
\end{figure}

For charmonium in the hadron gas, they mainly suffer the suppression by pion and rho mesons, with
reactions $J/\psi+\pi\leftrightarrow D+{\bar D}^{*},D^{*}+{\bar D}$ and
$J/\psi+\rho\leftrightarrow D^{*}+{\bar D}^{*},D+{\bar D}$.
Loss term in hadron gas phase $\alpha^{\mathrm{HG}}_{\Psi}$ is similar with that in
QGP. The degeneracies of pion and rho mesons are taken as $d_\pi=3$ and $d_\rho=9$, and
their mass are $m_{\pi}=0.135\ \mathrm{GeV}$ and $m_{\rho}=0.776\ \mathrm{GeV}$.
Both pion and rho mesons are taken as thermal distributions.
The inelastic cross-sections between $J/\psi$ and ($\pi$, $\rho$) is taken 
from~\cite{Grandchamp:2002wp, Lin:1999ad, Haglin:2000ar}. 
With above setup, the rate $\alpha^{\mathrm{HG}}_{\Psi}$ of charmonium suppression in hadron gas can be obtained. 
In order to better see the difference of charmonium suppression by QGP and hadron gas, 
lifetimes of different charmonium states 
are plotted in Fig.~\ref{figlife}.
Lifetime of $J/\psi$ is much larger in hadron gas compared with that in QGP. 
Here the geometry scale is used to get the rates of dissociation for 
excited states ($\chi_c,\psi^\prime$) in hadron gas. 
With increasing temperature, both densities of gluons and ($\pi,\rho$) mesons increase, 
which results in strong decreasement of charmonium lifetimes, see Fig.\ref{figlife}.

\begin{figure}[b]
  \includegraphics[width=0.9\linewidth]{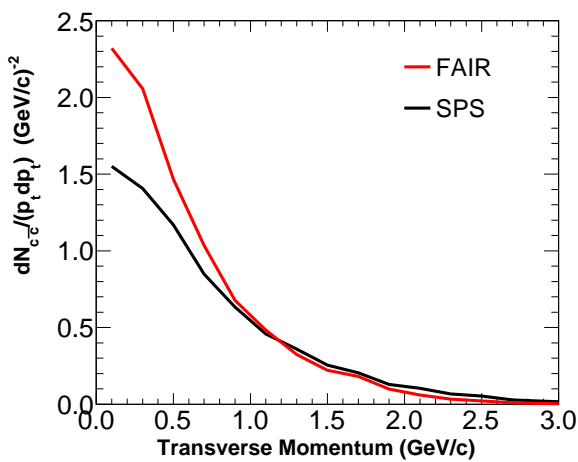}
  \caption{(Color Online) Normalized distribution of charm number as a function of transverse momentum
at FAIR and SPS simulated by PYTHIA. }
  \label{figCdist}
\end{figure}

There are also regenerated charmonium from recombination of $c\bar c$ or $D\bar D$ pairs, see the 
inverse reactions of charmonium dissociations. 
At these low colliding energies, the probability of producing more than one $c\bar c$ pairs in 
nucleus-nucleus collisions in one event is small. The regenerated charmonium is mainly 
from the same $c\bar c$ or $D\bar D$ pair. 
The momentum of produced $c$ and $\bar c$ in one pair have strong back-to-back correlation, which strongly suppress 
the probability of $c\bar c$ meeting again, and so the regeneration of charmonium. 
Considering that $c\bar c$ is hard to reach thermalization, we take the 
langevin equation for the evolutions of charm quark and D mesons, 
\begin{equation}
{d{\vec p}\over dt}=-\gamma (T){\vec p}+{\vec \eta}
\label{Eqlan}
\end{equation}
where $\gamma(T)=aT^2$, with $a=2\times 10^{-6}$ $\mathrm{(fm/c)^{-1}MeV^{-2}}$~\cite{Zhu:2007ne,Svetitsky:1996nj}.
${\vec \eta}$
is a Gaussian noise variable, and it can be obtained by the fluctuation-dissipation relation
in equilibrium~\cite{Svetitsky:1996nj}. 
The initial distribution of charm pairs in momentum space can be generated by PYTHIA. 
The total cross-section of $c\bar c$ is taken as $\sigma_{pp}^{c\bar c}/\sigma_{pp}^{J/\psi}=150$~\cite{Linnyk:2006ti}. And 
the normalized transverse momentum distribution of charm quark in pp collisions is given in Fig.\ref{figCdist}. 
As one can see,
at FAIR, more charm is distributed at low $p_t$ region compared with that at SPS. 
With momentum conservation, 
their initial momentum are always with the same magnitude but opposite direction, $\bf p_c=-\bf p_{\bar c}$. 
For the initial distribution of charm quark in 
coordinate space, it is randomly 
determined by Monte Carlo method, with the 
probability of $P_{c(\bar c)}({\bf r})$,
\begin{align}
P_{c(\bar c)}({\bf r})={T_A({\bf r}+{{\bf b}\over 2})T_B({\bf r}-{{\bf b}\over 2})\over 
\int T_A({\bf r}+{{\bf b}\over 2})T_B({\bf r}-{{\bf b}\over 2})d{\bf r}}
\end{align}
$c$ and $\bar c$ are produced at the same point due to their large mass. 

Once charm pairs are produced at the same point moving in the opposite direction, their 
evolution is described by the langevin equation (\ref{Eqlan}) untill freeze-out ($T_{\mathrm{freeze-out}}=0.12$ GeV) 
of the medium. During the 
evolution of $c$ and $\bar c$, they may recombine into a new charmonium. The probability depends on the distance 
between c and $\bar c$ and their relative momentum. It is taken as the conditions given in~\cite{yunpeng:2014}.
Once $c$ and $\bar c$ meet again during the evolution of medium, a new charmonium is formed and also survive from 
the hot medium (the following suppression of regenerated charmonium is neglected). 
After the evolution of hot medium, one can get the yields of charmonium 
from initial production by Boltzmann equation (\ref{tran}), and also regeneration from $c\bar c$ recombination 
by langevin equation (\ref{Eqlan}), and so the nuclear modification factor $R_{AA}$.  

\section*{V. Results and Analyse}
\begin{figure}[b]
  \includegraphics[width=0.9\linewidth]{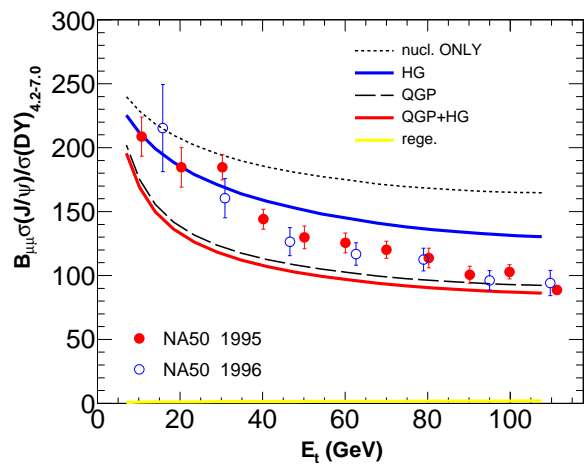}
  \caption{(Color Online) Charmonium suppression as a function of
transverse energy $E_t$ in $\sqrt{s_{NN}}=17.3$ GeV Pb+Pb collisions.
Dotted line only includes 
nucleon absorption. 
Black dashed line also includes QGP suppression, 
red solid line includes effects of both QGP and hadron gas. 
Blue solid line asumes no existence of QGP, 
and only includes contribution of hadron gas. All lines include cold nuclear matter effects. 
Yellow solid line is for the regeneartion of charmonium from langevin equation. 
Data is from~\cite{Abreu:2000xe,Topilskaya:2003iy}.
}
\label{figRAA_sps}
\end{figure}

With all the setup about initial charmonium and charm quark evolutions,
we give the charmonium nuclear modification factor $R_{AA}$ and average
transverse momentum square $\langle p_t^2\rangle$. In Fig.\ref{figRAA_sps}, 
Drell-Yan cross section is obtained with GRV-LO parton distribution function~\cite{Gluck:1991ng}. 
The regenerated charmonium from
$c\bar c$ recombination in QGP and $D\bar D$ recombination in hadron gas is neglectable (yellow solid line). 
During the time $\tau<\tau_0$, medium does not reach local equilibrium. 
The temperature of medium is taken as $T(x_t,\tau=\tau_0)$ 
for the charm evolution during 
this time. This may underestimate the effects of hot medium on charm thermalization. 
Drag coefficient is assumed to decrease with temperature. So the effects of hadron gas phase is much weaker than that 
of QGP. 
$c$ and $\bar c$ are hard 
to be thermalized in the hot medium, and moves almost in the opposite direction. 
This strongly suppress the probability of them meeting again 
during the evolution in the later time and the regeneration of charmonium. 
Nucleon absorption gives large suppression on charmonium yields, 
but is not enough for the charmonium suppression in all centralities. 
When the contribution of QGP is included, it can explain the data well in central collisions. 
For the excited states of charmonium which contribute 40\% to the final yield of total $J/\psi$, 
most of them are dissolved in QGP by color Debye screening, due to 
their small dissociation temperature $T_d\sim1.1T_c$. This makes nuclear modification factor of $J/\psi$ small 
in central collisions. And it overestimates the charmonium suppression in semi-central and peripheral collisions. 
When hadron gas is also included, the difference between black dashed and red solid lines is small (see Fig.\ref{figRAA_sps}). 
Because most of the excited states are already dissolved in QGP, and the lifetime of the left $J/\psi$ in hadron gas 
is much longer than that in QGP. This gives the small effect of hadron gas on final $J/\psi$ yields. 

The effects of QGP on charmonium is still an open question at SPS and FAIR. 
Considering that charmonium suppression
in nucleus-nucleus collisions is also caused by nucleon absorption and hadron gas, we should check
if these effects are enough for the data of charmonium suppression and the necessity of QGP.
We give the $J/\psi$ nuclear modification factor with the 
assumption that there is no QGP produced at SPS even when $T>T_c$ (blue solid line). This underestimates 
the charmonium suppression in central collisions, but explain the data well in peripheral collision. 
This is a good proof that there should be QGP produced in central 
collision and no QGP in peripheral collisions. 

\begin{figure}[b]
  \includegraphics[width=0.9\linewidth]{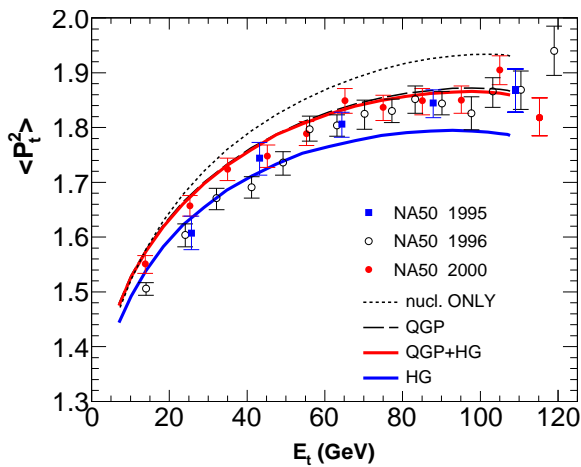}
  \caption{(Color Online) Averaged transverse momentum square of charmonium $\langle p_t^2\rangle$ as a function of
transverse energy $E_t$ (GeV) in $\sqrt{s_{NN}}=17.3$ GeV Pb+Pb collisions.
Dotted line only includes 
nucleon absorption. 
Black dashed line also includes effect of QGP, 
red solid line includes effects of both QGP and hadron gas. 
Blue solid line asumes no existence of QGP, 
and only includes contribution of hadron gas. All the lines include cold nuclear matter effects. 
Data is from~\cite{Abreu:2000xe,Topilskaya:2003iy}.
}
  \label{figpt2_sps}
\end{figure}

For the average transverse momentum square $\langle p_t^2\rangle$ in Fig.~\ref{figpt2_sps},
When only nucleon absorption is included, it gives larger $\langle p_t^2\rangle$ (black dotted line). 
The increasing tendency with $E_t$ is caused by Cronin effect, which is stronger in central collisions. 
It increase $\langle p_t^2\rangle$ of $J/\psi$ 
in pA and AA collisions. When QGP is 
included, charmonium suppression by hot medium is stronger in the center of fireball, where Cronin effect is also stronger. 
The dissociation of charmonium by QGP reduces 
the average transverse momentum square (black dashed line). It explains the data well at central collisions. 
After the evolution of QGP, 
effects of hadron gas at $T<T_c$ is neglectable for both $R_{AA}$ and $\langle p_t^2\rangle$ (red solid line). 
With the assumption of no phase transition and hot medium stays in hadron phase at all temperature, 
$\langle p_t^2\rangle$ can better explain the data well at peripheral collisions (blue solid line), just like $R_{AA}$. 

From the calculations of both $R_{AA}$ and $\langle p_t^2\rangle$, it seems that 
effect of QGP is important and necessary from these two observables in central collisions. But at peripheral 
collisions, effects of hadron gas and nucleon absorption explain the data well, and there should be no QGP 
produced. Charm quarks are hard to be thermalized at SPS and FAIR, and their back-to-back correlation of momentum 
strongly suppresses the regeneration of charmonium. 
Regeneration of charmonium should also be neglected 
at FAIR. 

\begin{figure}[b]
  \includegraphics[width=0.9\linewidth]{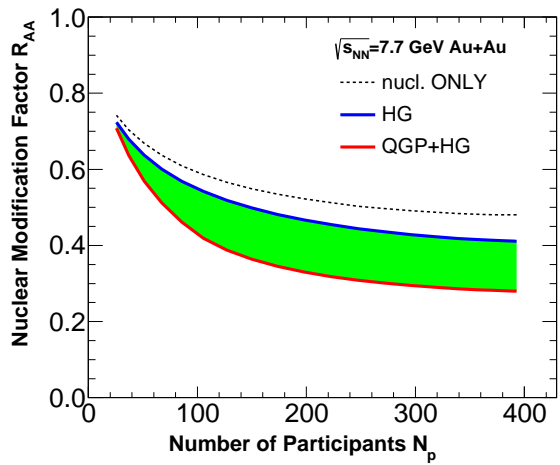}
  \caption{(Color Online) Charmonium nuclear modification factor $R_{AA}$ as a function of number of participants in 
$\sqrt{s_{NN}}=7.7$ GeV Au+Au collisions. 
Black dotted line includes only nuclear absorption, red solid line also includes the effects of QGP and hadron phase. 
Blue solid line assumes no existence of QGP, and charmonium only suffer suppression from nuclear absorption 
and hadron phase. The color band is caused by the uncertainty of QGP.  
}
  \label{figRAA_fair}
\end{figure}

Fixing the data at SPS, we give predictions at FAIR now.
For the parameters used at FAIR, we take the cross-section of nuclear absorption $\sigma^{abs}$ as 
$(4.5,4.5,7.9)$ mb for ($J/\psi,\chi_c,\psi^\prime$). The colliding energy of FAIR is close to that of SPS, 
so the parameter for Cronin effect is taken as the same of SPS. The 
average transverse momentum of $J/\psi$ produced at FAIR in pp collisions is $\langle p_t^2\rangle_{pp}=1.13\ \mathrm{(GeV/c)^2}$. 

In Fig.\ref{figRAA_fair}, charmonium suffer strong suppression from nuclear absorption when they are produced in the 
initial colliding times (black dotted line), just like at SPS. Considering that the energy density produced at FAIR 
is smaller than that at SPS, the charmonium suppression without QGP effects is given. It gives the upper limit of 
charmonium nuclear modification factor $R_{AA}$. But in the most central collisions, the 
temperature of the medium is higher than the critical temperature of phase transition, $T_c(\mu_B)$. There should be QGP 
produced in most central collisions, and we have to include color Debye screening on charmonium suppression (red solid line). 
Considering the uncertainty of QGP produced at these low colliding energies (SPS and FAIR),  
Two limits of $R_{AA}$ are given with and without the effects of QGP respectively. 
Comparing these limits with the experimental data helps us 
to better understand the realistic evolution of charmonium in heavy ion collisions. 
\begin{figure}[b]
  \includegraphics[width=0.9\linewidth]{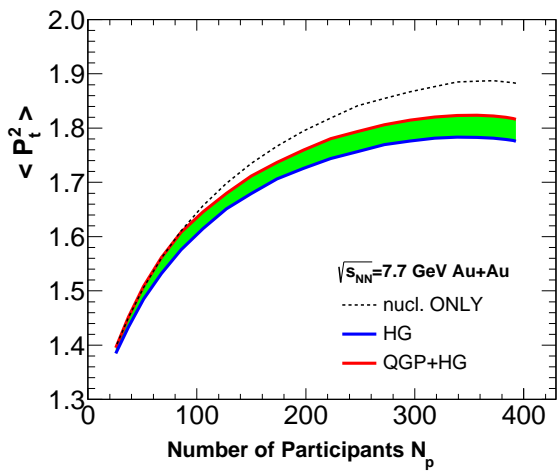}
  \caption{(Color Online) Charmonium average transverse momentum square $\langle p_t^2\rangle$
as a function of number of participants in $\sqrt{s_{NN}}=7.7$ GeV Au+Au collisions.
Black dotted line includes only nuclear absorption, red solid line also includes the effects of QGP and hadron phase. 
Blue solid line assumes no existence of QGP, and charmonium only suffer suppression from nuclear absorption 
and hadron phase. The color band is caused by the uncertainty of QGP.  
}
  \label{figpt2_fair}
\end{figure}

For the average transverse momentum square at FAIR, 
it shows similar tendency with the result of SPS. 
The increasement of black dotted line in Fig.\ref{figpt2_fair} is due to the cronin effect. 
This effect is proportional to the thickness of nucleus, and is stronger in central collisions. 
Since the matter produced in central collisions is 
denser and hotter than that in peripheral collisions, the 
$J/\psi$ has more chance to be dissolved by QGP, 
especially for excited states of charmonium. 
This suppresses the $\langle p_t^2\rangle$ after including the hot medium. That's why 
$\langle p_t^2\rangle $ with QGP (red solid line) become smaller 
compared with the situation with only nuclear absorption (black dotted line). 
With only hadron phase, the inelastic cross section between $J/\psi$ and ($\pi,\rho$) is taken as 
a constance when $s=(p_{J/\psi}+p_{(\pi,\rho)})^2$ is larger than a threshold. This results in a relatively small 
$\langle p_t^2\rangle$ (blue solid line) compared with the situation of QGP. 
The real situation of $R_{AA}$ and $\langle p_t^2\rangle$ should be inside the color bands in 
Fig.\ref{figRAA_fair} and Fig.\ref{figpt2_fair}. It should be 
close to the line with QGP (red solid line) at central collisions, and close 
to the line with only hadron phase (blue solid line) at peripheral collisions. 

\section*{VI. Conclusion}
In summary,
based on a realistic description of open and hidden heavy flavor evolution in hot medium, we calculate charmonium
suppression and production with and without the existence of QGP at SPS and FAIR. 
With only nuclear absorption and hadron phase, both charmoinum nuclear modification factor $R_{AA}$ and 
average transverse momentum square $\langle p_t^2\rangle$ explain the data well at peripheral collisions, 
but it underestimate the charmonium suppression in central collisions. 
With contributions of QGP, 
color Debye screening on excited charmonium 
states strongly suppress the final yield of $J/\psi$, and can explain both $R_{AA}$ and $\langle p_t^2\rangle$ well. 
From two limits given in Fig.\ref{figRAA_sps} and Fig.\ref{figpt2_sps}, one can come to a conclusion that QGP should be produced at 
the most central collisions and not produced at peripheral collisions at SPS. Charm and anti-charm quarks is hard to 
be thermalized at these colliding energies, and have small probability to recombine together during their evolution in the hot medium. 
The regeneration of 
charmonium is strongly suppressed. 
Based on above conclusions, predictions of charmonium observables are given at 
FAIR $\sqrt{s_{NN}}=7.7$ GeV Au+Au collisions. 

\appendix {\bf Acknowledgment}: Author thanks Dr.Pengfei Zhuang in Tsinghua University and Xiaojian Du in TAMU
for helpful discussions. 



\end{document}